\documentclass[
    ,final            
  ]
  {aipproc}

\layoutstyle{8x11double}
\usepackage{hyperref}

\def\be{\begin{equation}}
\def\ee{\end{equation}}
\def\bea{\begin{eqnarray}}
\def\eea{\end{eqnarray}}

\newcommand{\lsim}{\mbox{\raisebox{-.6ex}{~$\stackrel{<}{\sim}$~}}}

\def\sss{\scriptscriptstyle}
\def\vk{\mathbf{k}}
\def\mf{m_{\sss F}}
\def\nf{n_{\sss F}}

\def\gf{g_{\sss F}}
\def\EF{E_{\sss F}}
\def\pF{p_{\sss F}}

\begin{document}

\title[Cosmic ferromagnetism]{Magnetic domain walls of relic fermions  as Dark Energy}


\author{Urjit A. Yajnik}{
  address={Department of Physics, Indian Institute of Technology,
Bombay, Mumbai 400076 INDIA}, 
}

\copyrightyear  {2005}

\begin{abstract}
We show that relic fermions of the Big Bang can enter a ferromagnetic 
state if they possess a magnetic moment and satisfy the requirements of 
Stoner theory of itinerant ferromagnetism.  The domain walls of this  
ferromagnetism can successfully simulate Dark Energy over the observable
epoch spanning $\sim 10$ billion years. We obtain conditions on the anomalous  
magnetic moment of such fermions and their masses. Known neutrinos fail to
satisfy the requirements thus pointing to the possibility of a new ultralight
sector in Particle Physics.
\end{abstract}

\maketitle


\section{Introduction}
The recent strong evidence \cite{lamCDM} \cite{ParamWMAP} for the 
presence of a Cosmological Constant
in the Universe, contributing energy density $\approx (0.03 eV)^4$ (in units
$\hbar=c=1$), accounting for close to $70\%$ of its contents 
presents a new challenge to fundamental physics. The enigma of the discovery 
of such a small value, or indeed that of possible exact zero value for it
has been discussed in \cite{weinberg}. A less stringent alternative is
to explore the possibility of a substance capable of mimicking the equation
of state $p=w\rho$ with $w$ close to $-1$ at present epoch as appropriate to 
Cosmological Constant, but by contrast, capable of undergoing dynamical 
evolution. A large number of proposals have emerged along these lines, some 
attempting to connect the explanation to other enigmas of Cosmology, 
while some with a connection to theories beyond the Planck scale. In this 
paper we consider the possibility of a more convetional explanation 
based known phenomena. It has been argued in  \cite{Battye:1999eq}  
that the equation of state obeyed by this form of energy could be
well fitted by a network of domain walls which obey an effective 
equation of state $p=(-2/3)\rho$, and called Solid Dark Matter (SDM). 
This possibility has been further examined  in 
\cite{Conversi:2004pi}\cite{Friedland:2002qs} 
and shown to be still consistent with more recent data. 
In this paper we suggest that the domain walls could be of the
same kind that occur in ferromagnets. The domain wall complex obeys the Kibble 
law \cite{Kibble:1980mv},  its energy density contribution scaling as  $1/S(t)$,  
($S$ being the Friedmann-Robertson-Walker (FRW) scale factor)
thus providing a candidate SDM. 

Approaches to ferromagnetism based on the  well known  Heisenberg  
hamiltonian rely on interaction between spins localised in space at lattice sites. 
Such models are inadequate in explaining the large spontantaneous
ferromagnetism of iron, nickel and cobalt. A more fruitful approach 
is provided by Stoner theory \cite{Stoner} which concerns delocalised or 
itinerant electrons.  In this approach the cooperative phenomenon is 
visulaised in two parts, first fermionic correlations, equivalently
the Exclusion Principle causes a deficit of fermions of the same spin
in a given region of space. This enhanced averge separation  of electrons 
leads to reduction in energy since the screened coulomb interaction is
repulsive. In turn same orientation of the fermions is favoured energetically. 
A crucial ingredient of implementation of this ansatz is the excess interaction
energy per particle, and in turn the density of states at the Fermi surface.

In the cosmological setting, the 
density of states will simply be the number density of a free gas at the Fermi
surface. Further,  we assume  neutral fermions with  mutual 
interaction of purely magnetic origin, which is repulsive for same
spin fermions. 
With  these hypotheses we obtain a relation constraining the values of the 
anomalous magnetic  moment  and mass of the fermions. The suggested 
mass range is $\lsim 10^{-5}eV$, tantalisingly close to the small values suggested 
by neutrino oscillations \cite{NuMass}. However, the anomalous magnetic moment 
required is far in excess of the experimental and astrophysical bounds on 
magnetic moments of the three known neutrino species, thus ruling them out 
as viable candidates. Further, as discussed in conclusion, the gauge interaction 
proposed may not be  Electromagnetism. Thus we have an explanation 
of Dark Energy within the framework of known physical phenomena 
which leads to the hypothesis  of a new class of extremely light 
particles with large magnetic moment and  also the corresponding new gauge 
force.

\section{Cosmological setting}

The cosmological epoch we focus on is the one when the energy density
in Dark Energy became comparable to that in the form of non-relativistic 
matter, an epoch indicated to be about 7 billion years in the past.
Assuming the energy density of walls scales as $1/S(t)$ and using
the law $1/S^3(t)$ for the matter component, we can determine the 
time $t_1$  when the two contribute equally to the energy density. 
Using values of density fractions $\Omega_m\approx 0.3$, 
for matter and $\Omega_\Lambda\approx 0.7$ for the Dark Energy gives
$(S_1/S_0)^2=3/7$ where $0$ refers to current epoch.  Photon temperature 
at this epoch is $T_1=4.18 K= 5.0\times 10^{-4}eV$. 

Let us refer to the particle species generically as $F$ and assume that 
it was relativistic  at the time of nucleosynthesis,
ie $\mf < 0.1 MeV$. There are two possibilities for their abundance.
One is that like neutrinos their abundance is similar to that of
photons, $\nf(t_1)=$ $ (S_0/S_1)^3 120 cm^{-3}$ \cite{KolTur}. A less restrictive
possibility which we would like to exploit is that there may be excess
abundance of these particles, characterised by a factor $\Upsilon$
relative to photons. These possibilities together, in the units 
\,$\hbar=c=1$ become
\be
\nf(t_1)\approx 3.2 \times 10^{-12} \Upsilon (eV)^3
\label{eq:abund}
\ee
Excess abundance of a relativistic species conflicts with
nucleosynthesis unless it occurs only after the latter is complete. 
However, nontrivial $\Upsilon$ can arise from late deacy of a weakly 
interacting heavy particle. At present  epoch $t_0$  this abundance
is constrained by the requirement that 
the density fraction  $\Omega_F$ of a potential \textsl{hot} dark 
matter member must remain less than $0.003$ 
\cite{ParamWMAP} \cite{Lazarides}.  Using the value for 
$\rho_{crit}=3.6\times 10^{-6} (eV)^4$,
\be
\label{eq:omega}
\Omega_F\ =\ \frac{\mf \nf(t_1)}{\rho_{crit}}\ 
=\ 2.25 \times 10^{-7} \times \Upsilon \left( \frac{\mf}{eV} \right)  < 0.003
\ee
Thus if the species $F$ is saturating this bound, $\Upsilon(\mf/eV)\sim$ $10^4$.
These considerations apply to any one independent species. For 
more than one species participating in ferromagnetism, appropriate 
modifications can be incorporated. Finally, we use the expression
for the Fermi energy 
\be
\label{eq:EF}
\EF=
(\pF^2 + \mf^2)^{1/2} -\mf
\ee
which accords with the non-relativistic expression, and
where the zero-temperature Fermi momentum  is 
$\pF = (3\pi^2 \nf)^{1/3}$ for number density $\nf$.

\section{The Stoner criterion  and cosmic ferromagnetism}
We now turn to collective magnetic properties of this gas. We assume 
that individual particles have an effective intrinsic magnetic moment 
\be
\label{eq:mudef}
\mu_{\sss F} \equiv \gf\frac{e\,\hbar}{2\mf} = \gf\mu_{\sss B}\frac{m_e}{\mf}
\ee
where $m_e$ is the electron mass, $\mu_{\sss B}$ is the Bohr magneton,
and $\gf$ is the gyromagnetic ratio which must be entirely anomalous since
we are assuming the fermions to be neutral.  For 
neutrinos the radiatively induced magnetic moment is expected to be small 
\cite{Marciano:1977wx}, or $\mu_{\sss F}/\mu_{\sss B}<$ $10^{-15}$ 
as derived in \cite{Bell:2005kz} 
under certain reasonable assumptions. In a more general setting, $\gf$ 
can be order unity as in the case of the neutron. No such particle is 
expected from terrestrial experiments, however most of the Universe seems 
to be composed of  particles not suggested by any terrestrial experiments 
and there is no reason to forbid their existence. Also  most unified theories
such as $E_8\otimes E_8'$ and likewise the gauge mediated supersymmetry
breaking scheme invoke an unobserved gauge sector and this may be a 
manifestation of such a sector.

The Pauli paramagnetic susceptibility $\chi_{\sss P}$ of a spin gas is 
usually small. Large susceptibility and spontaneous magnetization 
arise according to the Stoner ansatz \cite{Stoner} 
if there is an additional shift in single particle energies, 
proportional to the  difference between
the spin up ($N_{\uparrow}$) and the spin down ($N_{\downarrow}$) populations. 
A parameter $I$
is introduced to incorporate this, the single-particle energy spectrum being
\be
E_{\uparrow, \, \downarrow}(\vk) = E(\vk) - I 
\frac{N_{\uparrow, \, \downarrow}}{N}
\ee
Using this it is shown \cite{Brout}\cite{IbLu}\cite{Irkhins} 
that the ferromagnetic  susceptibility is
\be
\chi\ =\ \frac{\chi_{\sss P}}{1 - I {\displaystyle 3 \over \displaystyle 4\EF}}\ 
\ee
The condition for spontaneous magnetization is negative $\chi$, which is
ensured provided the second term in the denominator dominates. A sufficient
condition  for the gas to be spontaneously magnetised at zero temperature  
is the Stoner criterion,
\be
\label{eq:stoner}
I\ >\ \frac{4\EF}{3}
\ee
Further, it can be shown that concordance with the Curie-Weiss law
suggests a critical temperature for the ferromagnetic phase transition 
as  $T_c=I /4$. Thus, if the zero temperature condition is satisfied and
temperature is less than $T_c$, ferromagnetic state is possible.

\begin{figure}
  \includegraphics
  [width=\columnwidth]{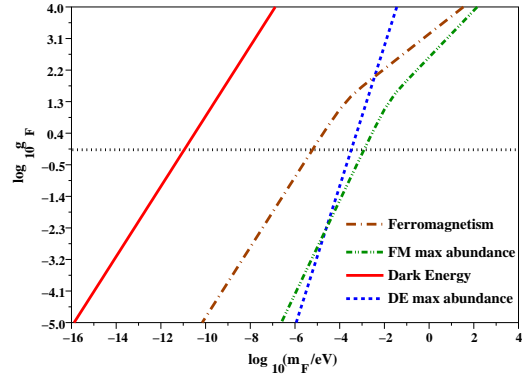}
  \caption{Permitted regions in parameter space $\log \mf$-$\log \gf$ are to
the left of the curves in all cases. Regions which satisfy
Stoner criteria are labelled Ferromagnetism (FM), without or with
maximum permissible abundance. The requirement of simulating Dark Energy 
(DE) restricts the regions further as shown. 
The horizontal dotted line corresponds to $\gf=1$.
}
\label{fig:plot}
\end{figure}

We now turn to the origin of the ferromagnetism.
Firstly we note that in the vicinity of a given fermion there is a deficiency of 
other fermions of same spin, the so called "exchange  hole" which is 
shown to exist in a standard derivation for which we refer the 
reader to \cite{FetWal}\cite{IbLu}. 
By averaging  over the Fermi sphere and integrating over the relative positions
this density deficiency can be estiamted to be
\be
\label{eq:deltan}
\Delta \nf\  =\   -0.86n_\nu
\ee
The upshot of this is that in principle a local population deficit of 
order unity is easy to obtain. Now consider a long range two particle
interaction $\gamma^2$ which is repulsive.  Compared to absence
of interaction and compared to a classical gas, the density deficit
causes a reduction in total energy density. This energy reduction
should be proportional to $\Delta \nf$. To retain the significance
of two particle interaction energy $\gamma^2$, we stipulate the relation
\be
I = \gamma^2 \frac{\vert \Delta \nf\vert}{\nf }
\label{eq:Idef}
\ee
We now make the assumption that for the fermions under consideration, this 
coupling arises from magnetic dipole-dipole interaction, which is repulsive 
between same spins.  The resulting increase in single particle
energy can be estimated  as (with $\mu_0$ the magnetic coupling in MKS units)  
\be
\label{eq:coupling}
\gamma^2\ =\  \kappa\mu_0 \mu_F^2 |\Delta \nf|
\ee 
being the mean field magnetic field
due to the deficit density of the magnetic moments,
times the magnetic moment $\mu_F$ of the single particle.
Here $\kappa$ is an unknown factor expected to be of order unity.
Note that the dipole interaction energy goes as inverse third power
of interparticle separation and hence correctly scales as $ |\Delta \nf|$.
Now we use (\ref{eq:mudef}), (\ref{eq:deltan}), (\ref{eq:Idef}), (\ref{eq:coupling}), and (\ref{eq:EF}) in
the Stoner criterion (\ref{eq:stoner}), and assume $|\Delta \nf|\approx \nf$, and $\kappa=1$ for simplicity.
The resulting regions in $\log\mf-\log\gf$ 
parameter space permitting ferromagnetic state to occur are shown 
in fig. \ref{fig:plot} for $\Upsilon=1$ and for the value of $\Upsilon$ which 
saturates the bound in eq. (\ref{eq:omega}). 
For $\gf\sim O(1)$, the allowed mass range of the species 
is $\leq 10^{-6}eV$ for $\Upsilon=1$, 
and relaxes to $\leq 10^{-3}eV$ for largest allowable $\Upsilon$.

The current upper bounds \cite{PDG} \cite{Raff90} on the neutrino magnetic 
moments are $10^{-10}$ for $\nu_e$ and $\nu_\mu$ and $4\times10^{-7}$ for $\nu_\tau$. 
These values are smaller by many orders of magnitude, $\sim (m_e/\mf)\times 10^{10}$  
( $\times 10^{7}$ for $\nu_\tau$) compared  to what is required here.
We must conclude that neutrinos cannot participate in such a mechanism. 
We also note that the excess abundance factor for neutrinos has been 
recently well constrained to be small, as deduced from neutrino oscillations
\cite{DHPPRS}\cite{AbaBeaBell}.
 
\section{Domain walls}
We now assume that at the phase transition when the ferromagnetic
state becomes favourable  a domain structure sets 
in due to finite corrrelations in the system. These domain walls are
not expected to be topologically stable. This is because the underlying
symmetry is $SU(2)$ of spin, which permits rotations within the vacuum
manifold for the defect to disentangle. However the situation is analogous
to the case of global internal symmetries  and the decay of the walls is
governed by tunneling processes as detailed later.  
We assume that the domain wall dynamics can be described by a 
Landau-Ginzburg effective lagrangian\cite{tobepub} for a vector order 
parameter $\bf S$ with a symmetry breaking self-interaction $\lambda 
({\bf S}\cdot{\bf S}-\sigma^2)^2$, where $\sigma$ determines the magnitude 
of the magnetization and can be related to $I$ and $\gamma^2$ introduced 
above.  From standard solitonic calculation 
\cite{Rajaraman} the domain walls have a width $w\sim(\sqrt{\lambda}\sigma)^{-1}$ 
and energy  per unit area $E/A \sim \sqrt{\lambda}\sigma^3$.

We can now estimate the energy trapped in the domain wall
structure and require that it must account for half 
of the total energy density of the Universe at the epoch $t_1$.
Let the 
domain wall structure be characterise by length scale $L$. Equivalently, 
there is one wall passing through a cubical volume of size $L^3$ on the average.
The energy density containd in such a wall is $E/wL^2$, while its average contribution
to the total density is $E/L^3$. If the walls are sufficiently distinguished as
a structure, we expect $w/L\ll 1$. 
\be
\rho_{walls}=\frac{E}{Aw}\frac{w}{L} \ge \frac{1}{2}\rho_{crit}\left( \frac{T_1}{T_0}\right)^4
\ee
For the generic case with no excess abundance this places a more stringent 
requirement on the allowed values of $\gf$ and $\mf$ as shown
in fig. \ref{fig:plot}, where $w/L=0.1$ is assumed. The restriction to smaller
mass values  enhances the magnetic energy stored in the walls, however the
mass values are in the range $\lsim 10^{-12}eV$, much smaller than known 
mass scales. If excess abundance is permitted this situation is considerably
relaxed. In this case a future problem would be to  understand the mechanism 
of the excess abundance.

One of the main sources of wall depletion is mutual collisions.
However the walls become non-relativistic soon after formation
and the free energy available for bulk motion reduces.
The walls can also spontaneously decay as was discussed in
\cite{PresVil92}. However the decay rate would be governed
by an exponential factor $\exp(-B/\lambda)$,
with $B$ the Euclidean action of the "bounce" \cite{Coleman}
of order unity. Stability of this complex for several billion years, 
compared to  intrinsic time scales of microscopic physics in the 
range $(0.03 eV)^{-1}$, requires the suppression factor 
to be $10^{-30}$ which is natural for $\lambda \sim 0.01$.

Finally we face the important question whether the gauge force responsible 
for this magnetism is Electromagnetism, $U(1)_{EM}$. Further investigation 
is required to determine the extent to which known phenomena such as \cite{Raff90} 
constrain the magnetic properties of such particles.  If magnetic moment 
of purely electromagnetic origin is too tightly constrained, the
gauge group may be a hidden sector $U(1)_H$.  Such a $U(1)_H$ would 
still mix  with the $U(1)_{EM}$. 
So long as $U(1)_{EM}$ is involved, an intriguing possibility is an 
explanation for the origin of the intergalactic magnetic fields \cite{Kulsrud}.
While the magnetic field averaged over the domains is zero, a 
deviation from the average, proportional to square root of the 
number of domains may suffice to provide the requisite seed \cite{tobepub}. 
An explanation for the $\gf$ value $\sim O(1)$ may be provided by
the existence of a hidden gauge group of the form 
$SU(N)_H\otimes U(1)_H$. In this case the situation may be similar to
the neutron, with  the fermion $F$ a strongly coupled
neutral bound state with large anomalous magnetic moment of the $U(1)_H$. 

A distinctive prediction of this scenario is that the Dark Energy
dominated era must end. As the Universe expands, the density
$\nf$ decreases and when the interaction stipulated in eq. (\ref{eq:coupling} )
becomes insignificant, spontaneous magnetism vanishes. The time scales are
expected to be comparable to cosmic time  as per the discussion
about stability of the walls. 
The disappearance of the domain walls would release some entropy.
This and other possible signatures of 
this scenario are model depedent and need further investigation.

If this scenario is correct then there is no fundamental Cosmological
Constant, returning General Relativity to its status where Einstein left it. 
Why the Higgs mechanism of electroweak theory induces no vacuum 
energy remains an open problem \cite{weinberg}.

 \begin{theacknowledgments}
 It is a pleasure to thank S. S. Jha for extensive discussion, Anjishnu
Sarkar for checking the calculations and C. P. Burgess  
and Mukund Rangamani for providing useful comments for the first version of 
the manuscript. This work is supported by a grant from Department 
of Science and Technology, India. 
\end{theacknowledgments}

\bibliographystyle{aipproc}

\end{document}